%% file: main.tex
\documentclass{article}

\input{preamble}

\title{ET-Flow: Equivariant Flow-Matching for Molecular Conformer Generation}

\author{%
    Majdi Hassan$^{*1}$ \\
    \And
    Nikhil Shenoy$^{*2,4}$ \\
    \And
    Jungyoon Lee$^{*1}$ \\
    \AND
    Hannes St\"ark$^{3}$ \\
    \And
    Stephan Thaler$^{4}$ \\
    \AND
    Dominique Beaini$^{1,4}$
    \AND  
    \textnormal{$^1$Mila \& Université de Montréal}\\
    \textnormal{$^2$University of British-Columbia}\\
    \textnormal{$^3$Massachusetts Institute of Technology}\\
    \textnormal{$^4$Valence Labs}\\
}
\begin{document}
% \nipsfinalcopy is no longer used

\nnfootnote{$^*$ Equal contribution.}
\nnfootnote{Correspondence to: \texttt{\{majdi.hassan,jungyoon.lee\}@umontreal.ca} and \texttt{nikhil@valencelabs.com}}

\maketitle
\begin{abstract}
Predicting low-energy molecular conformations given a molecular graph is an important but challenging task in computational drug discovery. Existing state-of-the-art approaches either resort to large scale transformer-based models that diffuse over conformer fields, or use computationally expensive methods to generate initial structures and diffuse over torsion angles. In this work, we introduce \textbf{E}quivariant \textbf{T}ransformer \textbf{Flow} ({\method}). We showcase that a well-designed flow matching approach with equivariance and harmonic prior alleviates the need for complex internal geometry calculations and large architectures, contrary to the prevailing methods in the field. Our approach results in a straightforward and scalable method that directly operates on all-atom coordinates with minimal assumptions. With the advantages of equivariance and flow matching, {\method} significantly increases the precision and physical validity of the generated conformers, while being a lighter model and faster at inference. Code is available \href{https://github.com/shenoynikhil/ETFlow}{https://github.com/shenoynikhil/ETFlow}.

\end{abstract}
\section{Introduction}
\begin{figure}
    \begin{subfigure}{0.77\textwidth}
        \centering
        \includegraphics[width=\textwidth, height=5.85 cm]{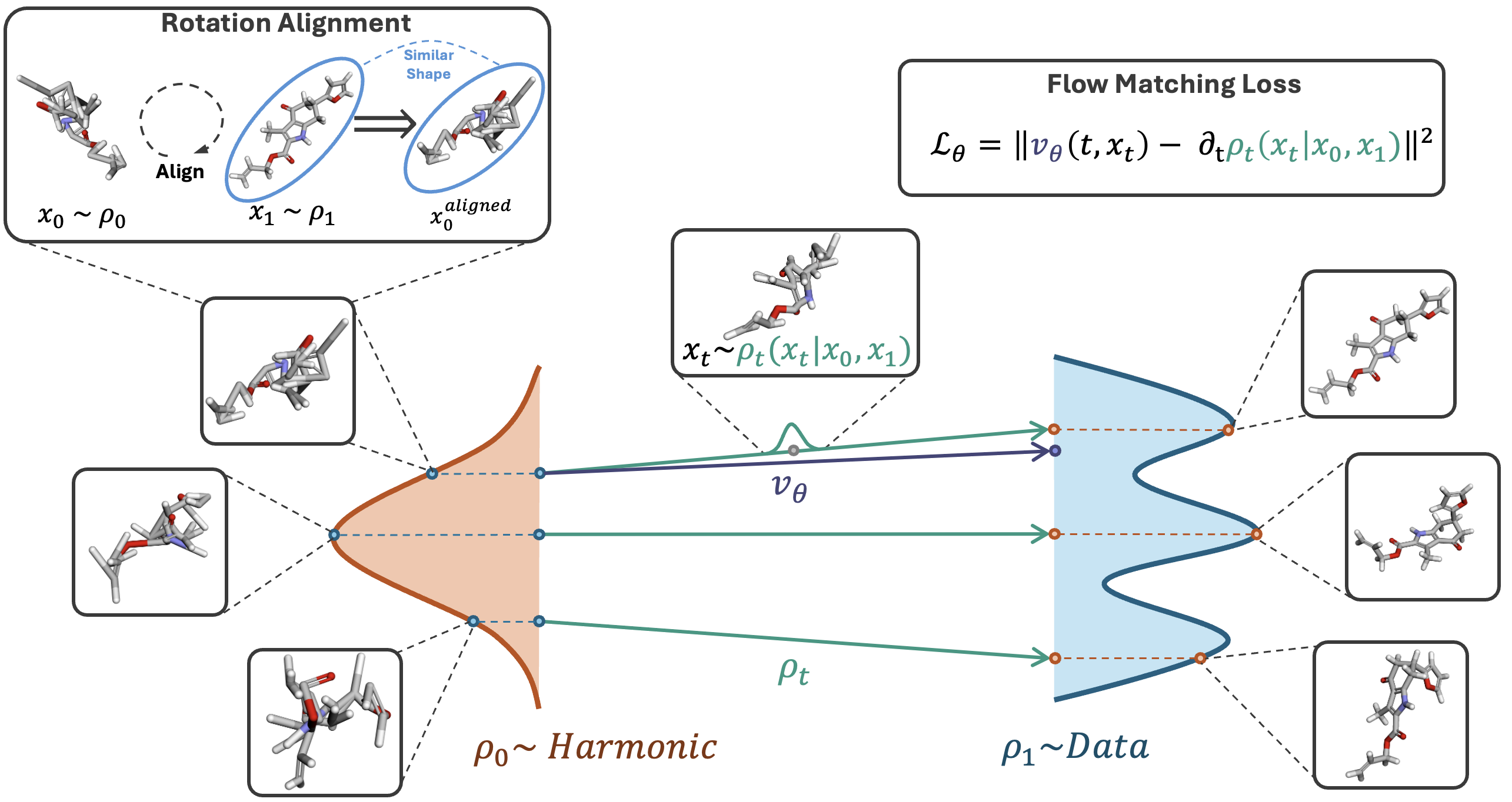}
        \caption{}
        \label{fig:etflow}
    \end{subfigure}
    \hfill
    \begin{subfigure}{0.22\textwidth}
        \centering
        \includegraphics[width=\textwidth, height=5.85 cm]{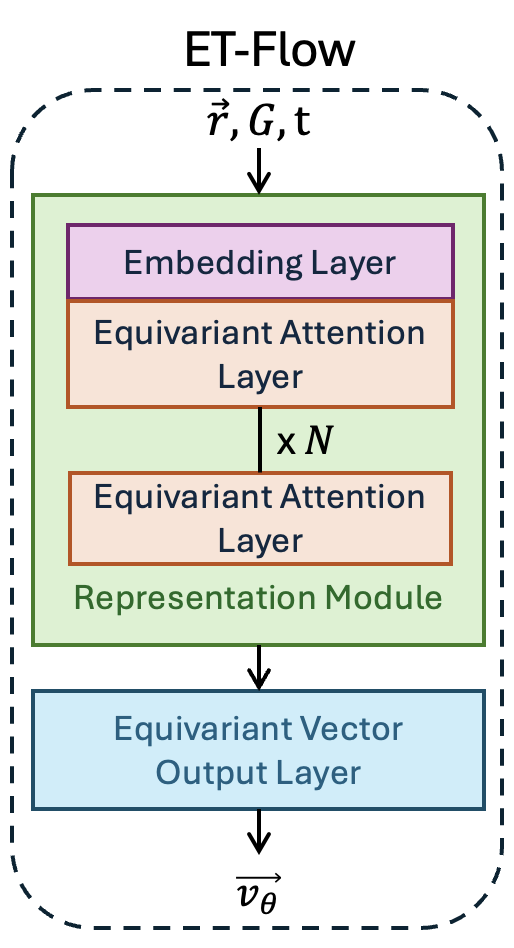}
        \caption{}
        \label{fig:etflow-arch}
    \end{subfigure}
     \caption{(a) Overview of {\method}. The model predicts a conditional vector field $\vec{v_\theta}$ using interpolated positions ($x_t$), molecular structure ($G$), and time-step ($t$). Samples are drawn from the harmonic prior ($x_0 \sim p_0$) and then rotationally aligned with the samples from data ($x_1 \sim p_1$). A conditional probability path is constructed between pairs of $x_0$ and $x_1$, and $x_t$ is then sampled from this path at a random time $t$. (b) The {\method} architecture consists of a representation module based on the TorchMD-NET architecture \citep{tholke2022torchmd} and an equivariant vector output module. For detailed architecture and input preprocessing information, see \autoref{app:architecture}.}
\end{figure}

Generating low-energy 3D representations of molecules, called \textit{conformers}, from the molecular graph is a fundamental task in computational chemistry as the 3D structure of a molecule is responsible for several biological, chemical and physical properties \citep{guimaraes2012use, schutt2018schnet, schutt2021equivariant, gasteiger2020directional, axelrod2023molecular}.
Conventional approaches to molecular conformer generation consist of stochastic and systematic methods. While stochastic methods such as Molecular Dynamics (MD) accurately generate conformations, they can be slow, cost-intensive, and have low sample diversity \citep{shim2011computational, ballard2015exploiting, de2016role, hawkins2017conformation, pracht2020automated}. Systematic (rule-based) methods \citep{hawkins2010conformer, bolton2011pubchem3d, li2007caesar, miteva2010frog2, cole2018knowledge, lagorce2009dg} that rely on torsional profiles and knowledge base of fragments are much faster but become less accurate with larger molecules. Therefore, there has been an increasing interest in developing scalable and accurate generative modeling methods in molecular conformer generation.

Existing machine learning based approaches use diffusion models \citep{ho2020denoising, song2019generative} to sample diverse and high quality samples given access to low-energy conformations. Prior methods typically fall into two categories: diffusing the atomic coordinates in the Cartesian space \citep{xu2022geodiff, wang2024swallowing} or diffusing along the internal geometry such as pairwise distances, bond angles, and torsion angles \citep{ganea2021geomol, jing2022torsional}. 

Early approaches based on diffusion \citep{shi2021learning, luo2021predicting, xu2022geodiff} faced challenges such as lengthy inference and training times as well as having lower accuracy compared to cheminformatics methods. Torsional Diffusion \citep{jing2022torsional} was the first to outperform cheminformatics methods by diffusing only on torsion angles after producing an initial conformer with the chemoinformatics tool RDKiT. This reliance on RDKiT structures instead of employing an end-to-end approach comes with several limitations, such as restricting the tool to applications where the local structures produced by RDKiT are of sufficient accuracy. Unlike prior approaches, the current state-of-the-art MCF \citep{wang2024swallowing} proposes a domain-agnostic approach by learning to diffuse over functions by scaling transformers and learning soft inductive bias from the data \citep{zhuang2022diffusion}. Consequently, it comes with drawbacks such as high computational demands due to large number of parameters, limited sample efficiency from a lack of inductive biases like euclidean symmetries, and potential difficulties in scenarios with sparse data — a common challenge in this field.

In this paper, we propose \textbf{E}quivariant \textbf{T}ransformer \textbf{Flow} ({\method}), a simple yet powerful flow-matching model designed to generate low-energy 3D structures of small molecules with minimal assumptions. We utilize flow matching \citep{lipman2022flow, albergo2023stochastic, liu2022flow}, which enables the learning of arbitrary probability paths beyond diffusion paths, enhancing both training and inference efficiency compared to conventional diffusion generative models. Departing from traditional equivariant architectures like EGNN \citep{satorras2021n}, we adopt an Equivariant Transformer \citep{tholke2022torchmd} to better capture geometric features. Additionally, our method integrates a Harmonic Prior \citep{jing2023eigenfold, stark2023harmonic}, leveraging the inductive bias that atoms connected by a bond should be in close proximity. We further optimize our flow matching objective by initially conducting rotational alignment on the harmonic prior, thereby constructing shorter probability paths between source and target distributions at minimal computational cost.

Our contributions can be summarized as follows:
\begin{enumerate}
    \item We obtain state-of-the-art precision for molecule conformer prediction, resulting in more physically realistic and reliable molecules for practitioners. We improve upon the previous methods by a large margin on ensemble property prediction.
    % \item We demonstrate the significance of carefully selected design choices such as incorporating equivariance and leveraging more effective priors in our simple yet well-engineered method. 
    \item We highlight the effectiveness of incorporating equivariance and more informed priors in generating physically-grounded molecules in our simple yet well-engineered method. 
    \item Our parameter-efficient model requires orders of magnitude fewer sampling steps than GeoDiff \citep{xu2022geodiff} and has  significantly fewer parameters than MCF \citep{wang2024swallowing}.
\end{enumerate}

\section{Background} \label{recent_work}
\textbf{Diffusion Generative Models.} Diffusion models \citep{song2019generative, song2020denoising, ho2020denoising} enables a high-quality and diverse sampling from an unknown data distribution by approximating the Stochastic Differential Equation(SDE) that maps a simple density i.e. Gaussian to the unknown data density. Concretely, it involves training a neural network to learn the score, represented as $\nabla_\x \log p_t(\x)$ of the diffused data. During inference, the model generates sample by iteratively solving the reverse SDE. However, diffusion models have inherent drawbacks, as they (i) require on longer training times (ii) are restricted to specific probability paths and (iii) depend on the use of complicated tricks to speed up sampling \citep{song2020denoising, zhang2022fast}.

\textbf{Flow Matching.} Flow Matching \citep{albergo2023stochastic, lipman2022flow, liu2022flow} provides a general framework to learn Continuous normalizing flows (CNFs) while improving upon diffusion models in simplicity, generality, and inference speed in several applications. Through simple regression against the vector field reminiscent of the score-matching objective in diffusion models, Flow matching has enabled a fast, simulation-free training of CNFs. Several subsequent studies have then expanded the scope of flow matching objective to manifolds \citep{chen2024flow}, arbitrary sources \citep{pooladian2023multisample}, and conditional flow matching with arbitrary transport maps and optimal couplings between source and target samples \citep{tong2023conditional}. 

\textbf{Molecular Conformer Generation.} Various machine learning (ML) based approaches \citep{kingma2013auto, liberti2014euclidean, dinh2016density, simm2019generative, shi2021learning, luo2021predicting, xu2021learning, ganea2021geomol, xu2022geodiff, jing2022torsional, wang2024swallowing} have been developed to improve upon the limitations of conventional methods, among which the most advanced are TorsionDiff \citep{jing2022torsional} and Molecular Conformer Fields (MCF) \citep{wang2024swallowing}. TorsionDiff designs a diffusion model on the torsion angles while incorporating the local structure from RDKiT ETKDG \citep{riniker2015better}. MCF trains a diffusion model over functions that map elements from the molecular graph to points in 3D space. 

\textbf{Equivariant Architectures for Atomistic Systems.} Inductive biases play an important role in generalization and sample efficiency. In the case of 3D atomistic modelling, one example of a useful inductive bias is the euclidean group $SO(3)$ which represents rotation equivariance in 3D space. Recently, various equivariant architectures \citep{duval2023hitchhiker} have been developed that act on both Cartesian \citep{satorras2021n, tholke2022torchmd, simeon2024tensornet, du2022se, frank2022so3krates} and spherical basis \citep{musaelian2023learning, batatia2022mace, fuchs2020se, liao2023equiformerv2, passaro2023reducing, anderson2019cormorant, thomas2018tensor}. For molecular conformer generation, initial methods like ConfGF, DGSM utilize invariant networks as they act upon inter-atomic distances, whereas the use of equivariant GNNs have been used in GeoDiff \citep{xu2022geodiff} and Torsional Diffusion \citep{jing2022torsional}. GeoDiff utilizes EGNN \citep{satorras2021n}, a Cartesian basis equivariant architecture while Torsional Diffusion uses Tensor Field Networks \citep{thomas2018tensor} to output pseudoscalars.

\section{Method}

We design {\method}, a scalable equivariant model that generates energy-minimized conformers given a molecular graph. In this section, we layout the framework to achieve this objective by detailing the generative process in flow matching, the rotation alignment between distributions, stochastic sampling, and finally the architecture details.

% \subsection{Preliminaries}
\textbf{Preliminaries} We define notation that we use throughout this paper. Inputs are continuous atom positions $\x \in \R^{N \times 3}$ where $N$ is the number of atoms. We use the notation $v_t(\x)$ interchangeably with $v(t, \x)$ for vector field.

\subsection{Flow Matching}\label{sec:flow-matching}

The aim is to learn a time-dependent vector field $v_t (x): \R^{N \times 3} \times [0, 1] \rightarrow \R^{N \times 3}$ associated with the transport map $X_t: \R^{N \times 3} \times [0, 1] \rightarrow \R^{N \times 3}$ that pushes forward samples from a base distribution $\rho_0$, often an easy-to-sample distribution, to samples from a more complex target distribution $\rho_1$, the low-energy conformations of a molecule. This can be defined as an ordinary differential equation (ODE),
\begin{equation}
    \dot{X_t}(\x) = v_t(X_t(\x)), \hspace{1 cm} X_{t=0} = \x_0,
\end{equation}
where $x_0 \sim \rho_0$. We can construct the $v_t$ via a time-differentiable interpolation between samples from $\rho_0$ and $\rho_1$ that gives rise to a probability path $\rho_t$ that we can easily sample \citep{lipman2022flow, liu2022flow, albergo2023building, tong2023conditional}. The general interpolation between samples $x_0 \sim \rho_0$ and $x_1 \sim \rho_1$ can be defined as:
\begin{equation}\label{eq:interpolent}
    I_t (\x_0, \x_1) = \alpha_t \x_1 + \beta_t \x_0.
\end{equation}
Given this interpolant that couples $\x_0$ and $\x_1$, we can define the conditional probability path as $\rho_t(\x | \x_0, \x_1) = \N(\x | I_t(\x_0, \x_1), \sigma_t^2\I)$, and the vector field can be computed as $v_t (\x) = \partial_t \rho_t(\x | \x_0, \x_1)$ which has the following form
\begin{equation}\label{eq:vector_field}
        v_t(\x) = \dot{\alpha}_t \x_1 + \dot{\beta}_t \x_0 + \dot{\sigma}_t \z \hspace{1 cm} \z \sim \N(0, \I).
\end{equation}
Here we use $\dot{\alpha}_t$ as a shorthand notation for $\partial_t\alpha_t$, and similarly we apply the same notation to $\beta$ and $\sigma$. In our work, we use linear interpolation where $\alpha_t = t$, $\beta_t = 1 - t$, and $\sigma_t = \sigma \sqrt{t(1 - t)}$, resulting in the vector field 
% \begin{equation}
%     x_t = t\x_1 - (1 - t)\x_0 + \sigma \sqrt{t(1 - t)} \z.
% \end{equation}
\begin{equation}
    v_t(\x) = \x_1 - \x_0 + \frac{1 - 2t}{2 \sqrt{t (1 - t)}} \z.
\end{equation}
Now, we can define the objective function for learning a vector field $v_\theta(\x)$ that generates a probability path $\rho_t$ between a base density $\rho_0$ and the target density $\rho_1$ as,
\begin{equation}\label{eq:loss}
    \mathcal{L} = \E_{t \sim \U(0, 1), \x \sim \rho_t(\x_0, \x_1)} \| v(t, \x) - v_\theta (t, \x) \|^2.
\end{equation}
For training, we sample (i) $\x_0 \sim \rho_0$, $\x_1 \sim \rho_1$, and $t \sim \U(0, 1)$, (ii) interpolate according to \autoref{eq:interpolent}, (iii) add noise from a standard Gaussian, and (iv) minimize the loss defined in \autoref{eq:loss}. For sampling, we sample $\x_0 \sim \rho_0$ and integrate from $t = 0$ to $t = 1$ using the Euler's method. At each time-step, the Euler solver iteratively predicts the vector field for $\x_t$ and updates its position $\x_{t + \Delta t} = \x_t + v_\theta(t, \x)\Delta t$. More details on the training and sampling algorithms are provided in \autoref{app:summary_procedures}.

\subsection{Alignment}\label{OT}

Several previous works \citep{tong2023conditional, klein2024equivariant, jing2024alphafold, song2024equivariant} demonstrate that constructing a straighter path between base distribution $\rho_0$ and target distribution $\rho_1$ minimizes the transport costs and improves performance. In our work, we reduce the transport costs between samples from the harmonic prior $\rho_0$ and samples from the data distribution $\rho_1$ by rotationally aligning them using the Kabsch algorithm \citep{Kabsch:a12999} similar to \citep{klein2024equivariant, jing2024alphafold}. This approach leads to faster convergence and reduces the path length between atoms by leveraging the similarity in "shape" of the samples as seen in \autoref{fig:etflow} without incurring high computational cost. %As a result, we observe a significant performance improvement relative to no-alignment shown in \autoref{tab:ablations}.

% Previous works showcase that optimal transport between samples from $\rho_0$ and $\rho_1$ demonstrate better performance since the paths are straighter (\citep{tong2023conditional}). 
% Furthermore, the alignment results in shorter, more direct vector fields between the corresponding atoms as seen in \autoref{fig:rot_align}. 

\subsection{Stochastic Sampling} 

We employ a variant of the stochastic sampling technique inspired by \citep{karras2022elucidating}. Specifically, we inject noise at each time step to construct an intermediate state, evaluate the vector field from the intermediate state, and then perform the deterministic ODE step from the noisy state. The original method utilizes a second-order integration, which averages the denoiser output at the noisy intermediate state and the state at the next time step after integration. 

\begin{wrapfigure}[18]{r}{0.35\textwidth}
    \centering
    \includegraphics[width=0.35\textwidth]{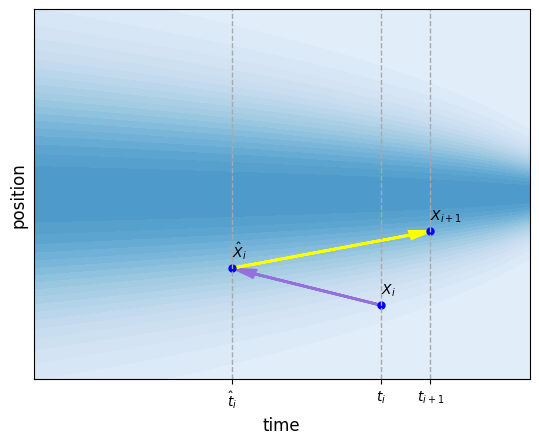}
    \caption{Stochastic sampling procedure used in inference. Noise is added to the positions $x_t$ indicated by the purple line, resulting in $\hat{x}_t$. Then, the model predicts the vector field $\hat{v}_t$ from $\hat{x}_t$ instead of $x_t$ indicted by the yellow line and updates $\hat{x}_t$ using $\hat{v}_t$ to get $x_{t+1}$.}
    \label{fig:stochastic_sampling}
\end{wrapfigure}
% \vspace{-0.5 cm}

In our experiment, we use the stochastic sampler without this second-order correction term, which empirically provided a performance boost comparable to the second-order method. We apply stochastic sampling only during the final part of the integration steps, specifically within the range $t \in [0.8, 1.0]$. This helps prevent drifting towards overpopulated density regions and improves the quality of the samples \citep{karras2022elucidating}. Stochastic sampling has improved both diversity and accuracy of the generated conformers, measured by Coverage and Average Minimum RMSD (AMR) respectively as shown in \autoref{tab:drugs}. Detailed information on the stochastic sampling algorithm is provided in \autoref{alg:sto-sampler}.

\subsection{Chirality Correction}\label{sec:chirality-correction}
While generating conformations, it is necessary to take account of the stereochemistry of atoms bonded to four distinct groups also referred to as tetrahedral chiral centers. To generate conformations with the correct chirality, we propose a simple \textit{post hoc} trick as done in GeoMol \citep{ganea2021geomol}. We compare the oriented volume (OV) (\autoref{eq:chiralitycorrection}) of the generated conformation and the required orientation with the RDKit tags. In the case of a mismatch, we simply flip the conformation against the z-axis. This correction step can be efficiently performed as a batched operation since it involves a simple comparison with the required RDKit tags and an inversion of position if necessary.

\begin{equation}\label{eq:chiralitycorrection}
    \text{OV}(\bm{p}_1, \bm{p}_2, \bm{p}_3, \bm{p}_4) = sign \left(\begin{vmatrix}
1   & 1   & 1   & 1  \\
x_1 & x_2 & x_3 & x_4\\
y_1 & y_2 & y_3 & y_4\\
z_1 & z_2 & z_3 & z_4\\
\end{vmatrix}\right).
\end{equation}

We also consider an alternative approach for chirality correction. Instead of using the \textit{post hoc} correction with our $O(3)$ equivariant architecture, we slightly tweak our architecture to make it $SO(3)$ equivariant by introducing a cross product term in the update layers. We compare these methods on both the GEOM-DRUGS and GEOM-QM9 dataset in \autoref{tab:drugs} and \autoref{tab:qm9}. Our base method ({\method}) corresponds to using the \textit{post hoc} correction whereas the $SO(3)$ variant is referred by {\method}-$SO(3)$. We empirically observe that using an additional chirality correction step is not only computationally efficient, but also performs better. We provide details on the architectural modification and proof of $SO(3)$ equivariance in \autoref{app:architecture} and \autoref{app:proof_of_so3} respectively.

\subsection{Architecture}\label{architecture}
% We utilize the equivariant transformer architecture as proposed in the TorchMD-NET \citep{tholke2022torchmd} which is designed using similar principles as the original Transformer \citep{vaswani2017attention}. {\method} (\autoref{fig:etflow-arch}) comprises of a (1) representation module based on the equivariant transformer architecture and (2) the equivariant vector output module. 

{\method} (\autoref{fig:etflow-arch}) consists of two main components: (1) a representation module based on the equivariant transformer architecture from TorchMD-NET \citep{tholke2022torchmd} and (2) the equivariant vector output module. In the representation module, an embedding layer encodes the inputs (atomic positions, atomic numbers, atom features, bond features and the time-step) into a set of invariant features. Initial equivariant features are constructed using normalized edge vectors where the edges are constructed using a radius graph of $10$ angstrom and the bonds from the 2D molecular graph. Then, a series of equivariant attention-based layers update both the invariant and equivariant features using a multi-head attention mechanism. Finally, the vector field is produced by the output layer, which updates the equivariant features using gated equivariant blocks \citep{schutt2018schnet}. Given that TorchMD-NET was originally designed for modeling neural network potentials, we implement several modifications to its architecture to better suit generative modeling, as detailed in \autoref{app:architecture}.

\section{Experiments}\label{sec:experiment}
We empirically evaluate {\method} by comparing the generated and ground-truth conformers in terms of distance-based RMSD (\autoref{sec:rmsd}) and chemical property based metrics (\autoref{sec:property}). We present the general experimental setups in \autoref{sec:setup}. The implementation details are provided in \autoref{app:implementation}.

\subsection{Experimental Setup}\label{sec:setup}
\textbf{Dataset}: We conduct our experiments on the GEOM dataset \citep{axelrod2022geom}, which offers curated conformer ensembles produced through meta-dynamics in CREST \citep{pracht2024crest}. Our primary focus is on GEOM-DRUGS, the most extensive and pharmacologically relevant subset comprising 304k drug-like molecules, each with an average of 44 atoms. We use a train/validation/test ($243473$/$30433$/$1000$) split as provided in \citep{ganea2021geomol} Additionally, we train and test model on GEOM-QM9, a subset of smaller molecules with an average of 11 atoms. Finally, in order to assess the model’s ability to generalize to larger molecules, we evaluate the model trained on GEOM-DRUGS on a GEOM-XL dataset, a subset of large  molecules with more than 100 atoms. The results for GEOM-QM9 and GEOM-XL can be found in the \autoref{app:additional_results}.

\textbf{Evaluation}: Our evaluation methodology is similar to that of \citep{jing2022torsional}. First, we look at RMSD based metrics like Coverage and Average Minimum RMSD (AMR) between generated and ground truth conformer ensembles. For this, we generate $2K$ conformers for a molecule with $K$ ground truth conformers. Second, we look at chemical similarity using properties like Energy ($E$), dipole moment ($\mu$), HOMO-LUMO gap ($\Delta \epsilon$) and the minimum energy ($E_{\min}$) calculated using xTB \citep{gfnxtb}.

\textbf{Baselines}: We benchmark {\method} against leading approaches outlined in Section \ref{recent_work}. Specifically, we assess the performance of GeoMol \citep{ganea2021geomol}, GeoDiff \citep{xu2022geodiff}, Torsional Diffusion \citep{jing2022torsional}, and MCF \citep{wang2024swallowing}. Notably, the most recent among these, MCF, has demonstrated superior performance across evaluation metrics compared to its predecessors. It's worth mentioning that GeoDiff initially utilized a limited subset of the GEOM-DRUGS dataset; thus, for a fair comparison, we consider its re-evaluated performance as presented in \citep{jing2022torsional}. 

% \subsection{Optimal Transport Alignment}\label{OT-exp}
% We introduce a preprocessing procedure termed optimal transport alignment. Simply put, for the training dataset exclusively, we adjust our sampled harmonic prior to each corresponding ground truth conformer using a method described in Section \autoref{OT}, aiming to straighten the probability path between the sampled and target distributions as much as possible. Here, we demonstrate the effect of this approximate OT alignment by examining the average path length of our vector field in Table \inshere. 
% \mh{Run an experiment that compares the average path length with alignment vs without alignment and provide results in small table}

\subsection{Ensemble RMSD}\label{sec:rmsd}
\begin{table*}[t]
\caption{Molecule conformer generation results on GEOM-DRUGS ($\delta = 0.75\text{\AA}$). {\method} - SS is {\method} with stochastic sampling and {\method} - $SO(3)$ is {\method} using the $SO(3)$ architecture for chirality correction. For {\method}, {\method}-SS and {\method}-$SO(3)$, we sample conformations over $50$ time-steps.}
\label{tab:drugs}
\centering
\begin{adjustbox}{max width=\textwidth}
\renewcommand{\arraystretch}{1.1} % Default value: 1
\begin{tabular}{lcccccccc}
\toprule
% \Xhline{3\arrayrulewidth}
 & \multicolumn{4}{c}{Recall} & \multicolumn{4}{c}{Precision} \\
% \Xhline{2\arrayrulewidth}
 & \multicolumn{2}{c}{Coverage $\uparrow$} & \multicolumn{2}{c}{AMR $\downarrow$} & \multicolumn{2}{c}{Coverage $\uparrow$} & \multicolumn{2}{c}{AMR $\downarrow$} \\
    \hline
 & mean & median & mean & median & mean & median & mean & median\\
    \hline
    GeoDiff         & 42.10 & 37.80 & 0.835 & 0.809 & 24.90 & 14.50 & 1.136 & 1.090 \\
    GeoMol          & 44.60 & 41.40 & 0.875 & 0.834 & 43.00 & 36.40 & 0.928 & 0.841 \\
    Torsional Diff. & 72.70 & 80.00 & 0.582 & 0.565 & 55.20 & 56.90 & 0.778 & 0.729 \\
    MCF - S (13M)          & 79.4 & 87.5 & 0.512 & 0.492 & 57.4 & 57.6 & 0.761 & 0.715 \\
    MCF - B (62M)         & 84.0 & 91.5 & 0.427 & 0.402 & 64.0 & 66.2 & 0.667 & 0.605 \\
    MCF - L (242M)          & \textbf{84.7} & \textbf{92.2} & \textbf{0.390} & \textbf{0.247} & 66.8 & 71.3 & 0.618 & 0.530 \\
    \hline
    {\method} (8.3M)       &  79.53 & 84.57 & 0.452 & 0.419 & 74.38 & 81.04 & 0.541 & 0.470 \\%post-hoc correction 
    {\method} - SS (8.3M)  &  79.62 & 84.63 & 0.439 & 0.406 & \textbf{75.19} & \textbf{81.66} & \textbf{0.517} & \textbf{0.442} \\%post-hoc correction
    % so3 correction
    {\method} - $SO(3)$ (9.1M)  & 78.18 & 83.33 & 0.480 & 0.459 & 67.27 & 71.15 & 0.637 & 0.567 \\%post-hoc correction
\midrule
\end{tabular}
\end{adjustbox}
\end{table*}

\begin{table*}[t]
\caption{Molecule conformer generation results on GEOM-QM9 ($\delta = 0.5\text{\AA}$). {\method} - $SO(3)$ is {\method} using the $SO(3)$ architecture for chirality correction. For both {\method} and {\method}-$SO(3)$, we sample conformations over $50$ time-steps.}
\label{tab:qm9}
\centering
\renewcommand{\arraystretch}{1.1} % Default value: 1
\begin{tabular}{lcccccccc}
% \Xhline{3\arrayrulewidth}
\toprule
 & \multicolumn{4}{c}{Recall} & \multicolumn{4}{c}{Precision} \\
 \Xhline{2\arrayrulewidth}
 & \multicolumn{2}{c}{Coverage $\uparrow$} & \multicolumn{2}{c}{AMR $\downarrow$} & \multicolumn{2}{c}{Coverage $\uparrow$} & \multicolumn{2}{c}{AMR $\downarrow$} \\
    \hline
    & mean & median & mean & median & mean & median & mean & median\\
    \hline
    CGCF            & 69.47 & 96.15           & 0.425 & 0.374 & 38.20 & 33.33           & 0.711 & 0.695 \\
    GeoDiff         & 76.50 & \textbf{100.00} & 0.297 & 0.229 & 50.00 & 33.50           & 1.524 & 0.510 \\
    GeoMol          & 91.50 & \textbf{100.00} & 0.225 & 0.193 & 87.60 & \textbf{100.00} & 0.270 & 0.241 \\
    Torsional Diff. & 92.80 & \textbf{100.00} & 0.178 & 0.147 & 92.70 & \textbf{100.00} & 0.221 & 0.195 \\
    MCF             & 95.0 & \textbf{100.00} & 0.103 & 0.044 & 93.7 & \textbf{100.00} & 0.119 & 0.055 \\
    \hline
    {\method} &  \textbf{96.47} & \textbf{100.00} & \textbf{0.073} & 0.047 & \textbf{94.05} & \textbf{100.00} & \textbf{0.098} & \textbf{0.039} \\%post-hoc correction
    {\method} - $SO(3)$ &  95.98 & \textbf{100.00} & 0.076 & \textbf{0.030} & 92.10 & \textbf{100.00} & 0.110 & 0.047 \\%post-hoc correction    
\Xhline{3\arrayrulewidth}\\
\end{tabular}
% 95.98	100.00	0.076	0.030	94.05	100.00	0.098	0.039
% 96.47	100.00	0.073	0.047	92.10	100.00	0.110	0.047
\end{table*}

As shown in \autoref{tab:drugs} and \autoref{tab:qm9}, {\method} outperforms all preceding methodologies and demonstrates competitive performance with the previous state-of-the-art, MCF \citep{wang2024swallowing}. Despite being significantly smaller with only 8.3M parameters, {\method} shows a substantial improvement in the quality of generated conformers, as evidenced by superior Precision metrics across all MCF models, including the largest MCF-L. When compared to MCF-S, which is closer in size, {\method} achieves markedly better Precision while the impact on Recall is less significant and limited to Recall Coverage. Notably, our Recall AMR remains competitive with much bigger MCF-B, underscoring the inherent advantage of our method in accurately predicting overall structures.

%\yo{optimistic version:}
% {\method} outperforms all previous methods on GEOM-DRUGS (as shown in \autoref{tab:drugs} and Figure \inshere), with a reduction in the recall AMR by \inshere\% and in the precision AMR by \inshere\% compared to the prior state-of-the-art approach. Our method demonstrates particular strength in precision metrics, signaling a significant enhancement in the quality of generated conformers. Unlike Torsional Diffusion, which achieved denoising steps as low as 20-50 through reduced degrees of freedom and a strong RDKit prior, our approach avoids limiting the diversity of generated samples as indicated by recall metrics, while keeping the low number of denoising steps. \yo{Also add comments between MCF vs. ours} This capability stems from our method leveraging a stronger prior, a more optimal probability path, and a flexible generative approach, enabling the generation of diverse, high-quality samples.

\subsection{Coverage Threshold 
Plots}\label{sec:threshold_plots}
\begin{figure}[h]
    \centering
    \includegraphics[width=\linewidth]{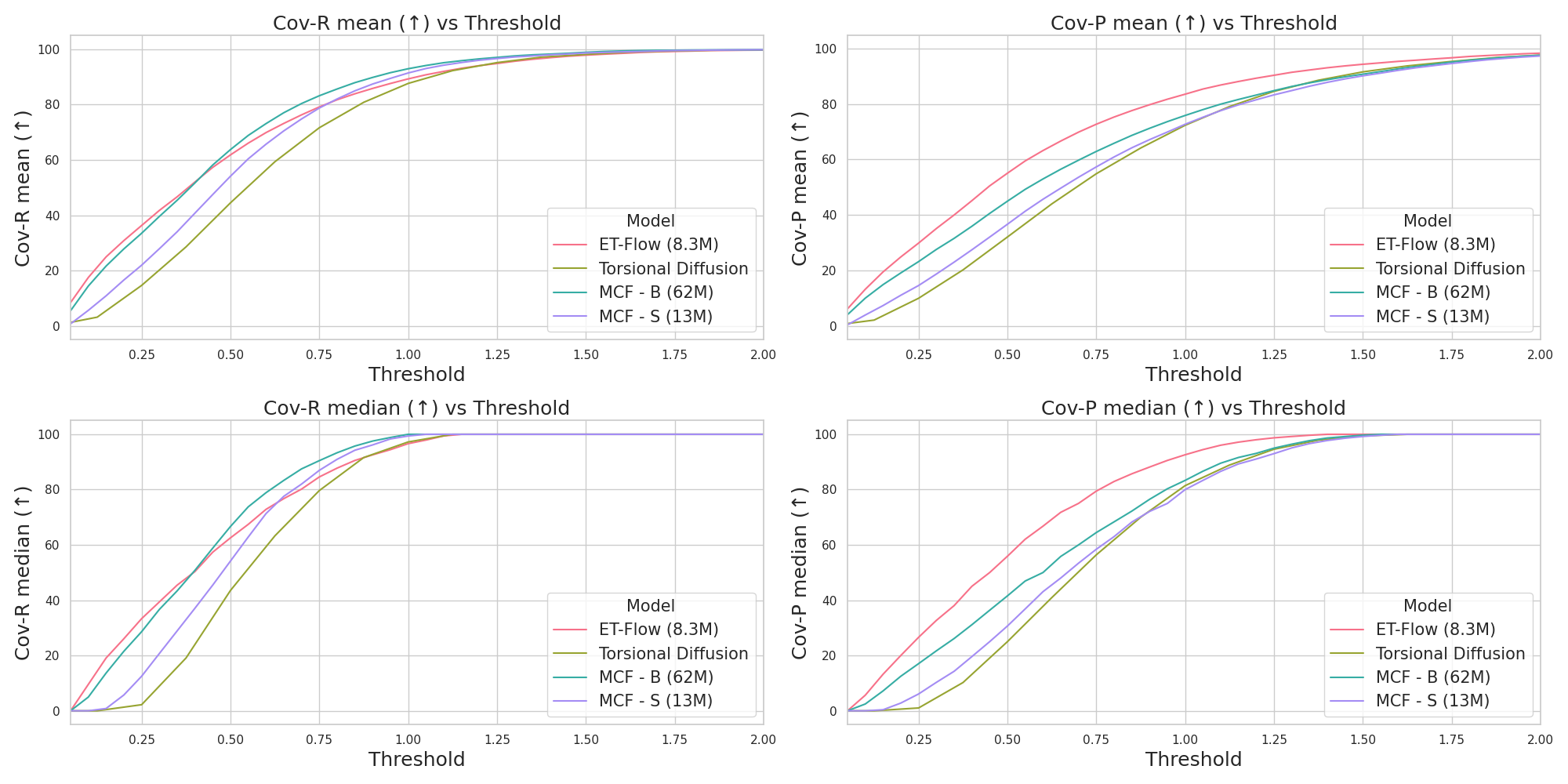}
    \caption{Recall and Precision Coverage result on GEOM-DRUGS as a function of the threshold distance. {\method} outperforms TorsionDiff by a large margin especially in a lower threshold region. We emphasize the better performance of {\method} at lower thresholds in both Recall and Precision metrics.}
    \label{fig:cov_mean}
\end{figure}

We compare the coverage metrics of {\method} against Torsional diffusion \citep{jing2022torsional} and MCF \citep{wang2024swallowing} against a wide range of thresholds on the GEOM DRUGS dataset in \autoref{fig:cov_mean}.
{\method} consistently outperforms previous methods in precision-based metrics. In terms of recall, our approach demonstrates better performance than Torsional Diffusion across all thresholds. Despite MCF performing better at higher thresholds, {\method} outperforms in the lower thresholds, underscoring its proficiency in generating accurate conformer predictions.

\subsection{Ensemble Properties}\label{sec:property}
RMSD provides a geometric measure for assessing ensemble quality, but it is also essential to consider the chemical similarity between generated and ground truth ensembles. For a random 100-molecule subset of the test set of GEOM-DRUGS, if a molecule has $K$ ground truth conformers, we generate a minimum of $2K$ and a maximum of 32 conformers per molecule. These conformers are then relaxed using GFN2-xTB \citep{gfnxtb}, and the Boltzmann-weighted properties of the generated and ground truth ensembles are compared. Specifically, using xTB \citep{gfnxtb}, we compute properties such as energy ($E$), dipole moment ($\mu$), HOMO-LUMO gap ($\Delta \epsilon$), and the minimum energy ($E_{min}$). \autoref{tab:ensemble} illustrates the median errors for {\method} and the baselines, highlighting our method's capability to produce chemically accurate ensembles. Notably, we achieve significant improvements over both TorsionDiff and MCF across all evaluated properties.

\begin{table}[ht]
\centering
\caption{Median averaged errors of ensemble properties between sampled and generated conformers ($E$, $\Delta \varepsilon$, $E_{min}$ in kcal/mol, and $\mu$ in debye).}\label{tab:ensemble}
\begin{tabular}{lcccc}
\toprule
& $E$ & $\mu$ & $\Delta \epsilon$ & $E_{\text{min}}$ \\
\midrule
OMEGA & 0.68 & 0.66 & 0.68 & 0.69 \\
GeoDiff & 0.31 & 0.35 & 0.89 & 0.39 \\
GeoMol & 0.42 & 0.34 & 0.59 & 0.40 \\
Torsional Diff. & 0.22 & 0.35 & 0.54 & 0.13 \\
MCF & 0.68$\pm$0.06 & 0.28$\pm$ 0.05 & 0.63$\pm$0.05 & 0.04$\pm$0.00\\
\midrule
% {\method}  & \textbf{0.19}$\pm$\textbf{0.02} & \textbf{0.19}$\pm$\textbf{0.04} & \textbf{0.36}$\pm$\textbf{0.04} & \textbf{0.02}$\pm$\textbf{0.01} \\
{\method}  & \textbf{0.18}$\pm$\textbf{0.01} & \textbf{0.18}$\pm$\textbf{0.01} & \textbf{0.35}$\pm$\textbf{0.06} & \textbf{0.02}$\pm$\textbf{0.00} \\
\bottomrule
\end{tabular}
\end{table}

\subsection{Inference Steps Ablation}\label{sec:inference-steps}
In \autoref{tab:drugs}, our sampling process with {\method} utilizes 50 inference steps. To evaluate the method's performance under constrained computational resources, we conducted an ablation study by progressively reducing the number of inference steps. Specifically, we sample for $5$, $10$ and $20$ time-steps. The results on GEOM-DRUGS are presented in \autoref{tab:steps-ablation}. We observed minimal performance degradation with a decrease in the number of steps. Notably, {\method} demonstrates high efficiency, maintaining performance across all precision and recall metrics even with as few as 5 inference steps. Interestingly, {\method} with 5 steps still achieves superior precision metrics compared to all existing methods. This underscores {\method}'s ability to generate high-quality conformations while operating within limited computational budgets.
\begin{table*}[t]
\caption{Ablation over number of inference steps on GEOM-DRUGS ($\delta = 0.75\text{\AA}$). Performance of {\method} at $5$ steps is competent across all metrics while also retaining state-of-the-art performance on precision metrics when compared with previous methods.}
\label{tab:steps-ablation}
\centering
\renewcommand{\arraystretch}{1.1} % Default value: 1
\begin{tabular}{lcccccccc}
% \Xhline{3\arrayrulewidth}
\toprule
 & \multicolumn{4}{c}{Recall} & \multicolumn{4}{c}{Precision} \\
 \midrule
 & \multicolumn{2}{c}{Coverage $\uparrow$} & \multicolumn{2}{c}{AMR $\downarrow$} & \multicolumn{2}{c}{Coverage $\uparrow$} & \multicolumn{2}{c}{AMR $\downarrow$} \\
    \hline
    & mean & median & mean & median & mean & median & mean & median\\
    \hline
    {\method} (5 Steps) & 77.84 & 82.21 & 0.476 & 0.443 & 74.03 & 80.8 & 0.55 & 0.474 \\
    {\method} (10 Steps) & 79.05 & 84.00 & 0.451 & 0.415 & 74.64 & \textbf{81.38} & 0.533 & 0.457 \\
    {\method} (20 Steps) & 79.29 & 84.04 & \textbf{0.449} & \textbf{0.413} & \textbf{74.89} & 81.32 & \textbf{0.531} & \textbf{0.454} \\
    {\method} (50 Steps) &  \textbf{79.53} & \textbf{84.57} & 0.452 & 0.419 & 74.38 & 81.04 & 0.541 & 0.470 \\
\bottomrule\\
\end{tabular}
\end{table*}

\subsection{Sampling Efficiency}\label{sec:inference_time}
\begin{figure}[h]
    \centering
    \includegraphics[width=\linewidth]{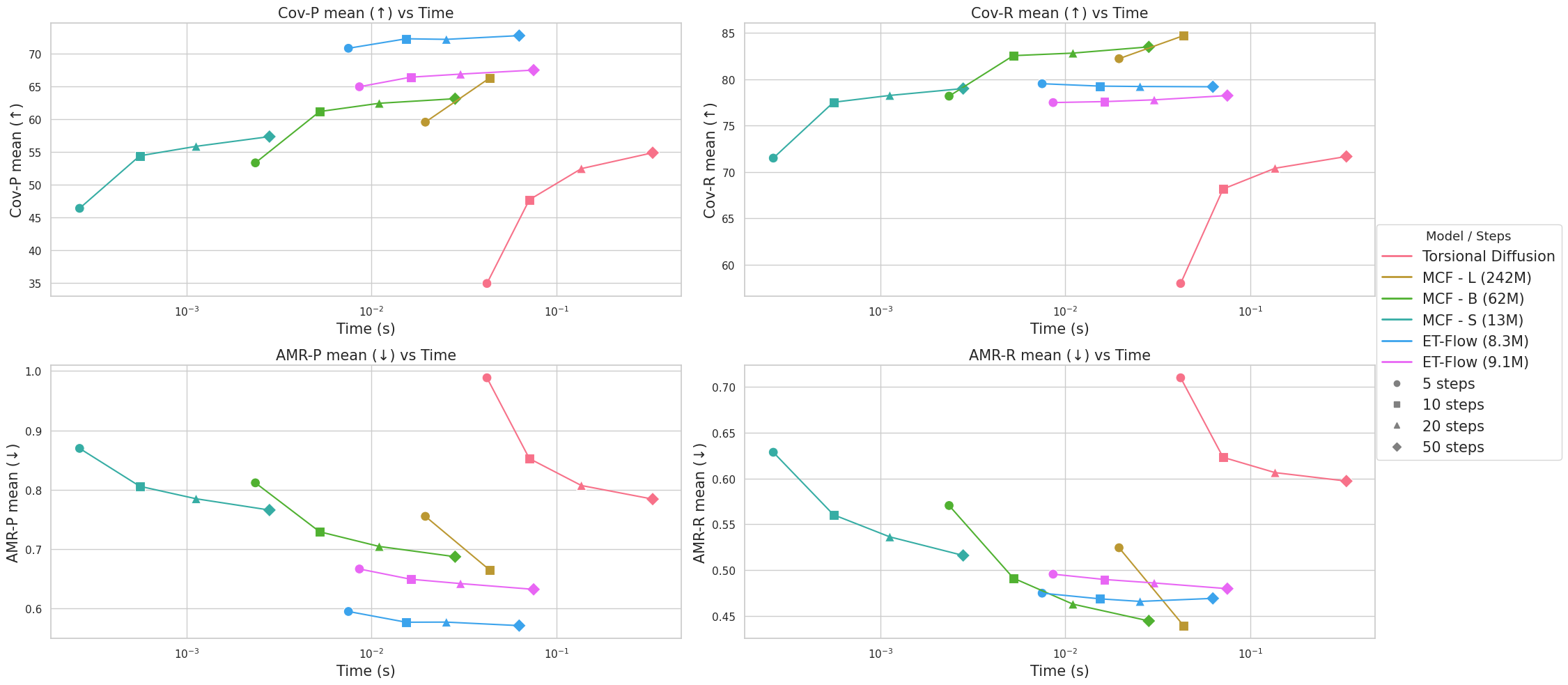}
    \caption{Sampling efficiency as a measure of the quality of Inference time with respect to the number of time steps on GEOM-DRUGS.} %{\method} outperforms TorsionDiff by a large margin.}
    \label{fig:inference_time}
\end{figure}

We demonstrate the ability of {\method} to generate samples efficiently. We evaluate the inference time per molecule over varying number of time steps and report the average time across 1000 random samples from the test set of GEOM-DRUGS. \autoref{fig:inference_time} shows that {\method} outperforms Torsional diffusion \citep{jing2022torsional} in inference across all time steps. While {\method} may not achieve the fastest raw inference times (potentially due to MCF variants benefiting from optimized CUDA kernels for attention), it maintains competitive speeds while ensuring higher precision. We suspect that concurrent work on improving equivariant operations with optimized CUDA kernels \citep{lee2024deconstructing} should lead to similar efficiency gains as seen in transformer-based architectures.

{\method} effectively balances performance and speed, making it ideal for tasks that require high sample quality with efficient computation. With the ability to generate high-quality samples in fewer time steps, e.g., 5 time steps, as indicated in \autoref{tab:steps-ablation}, {\method} is well-suited for scenarios demanding a large number of samples, as fewer steps lead to lower inference time per molecule. 
Additionally, we encountered difficulties running MCF-L for 20 and 50 steps, so those results have not been included. In summary, {\method} demonstrates efficient sampling, balancing precision and speed, making it highly effective for generating high-quality molecular samples while remaining competitive in inference time. 

% Combined with the results from , {\method} can effectively generate high quality samples with few time steps i.e. 5 time steps. This capability is especially useful for generating a large number of samples. \mh{Insert something about tradeoffs comparing mcf with etflow.}

% \subsection{\textit{post hoc} vs $SO(3)$ Chirality Correction}

% We conduct an ablation to understand if our correction method 

\section{Conclusion}
% \mh{
% Doms feedback: Pretty bad conclusion. Why do people care about this work??

% Because it's FAST
% Because it's PRECISE
% Because it's Physically accurate
% Because it generalizes well!

% Add a paragraph highlighting why your work is important and people should be using it.
% }

In this paper, we present our simple and scalable method {\method}, which utilizes an equivariant transformer with flow matching to achieve state-of-the-art performance on multiple molecular conformer generation benchmarks. By incorporating inductive biases, such as equivariance, and enhancing probability paths with a harmonic prior and RMSD alignment, we significantly improve the precision of the generated molecules, and consequently generate more physically plausible molecules. Importantly, our approach maintains parameter and speed efficiency, making it not only effective but also accessible for practical high-throughput applications. 

\section{Limitations And Future Works}\label{sec:future-works}

While {\method} demonstrates competitive performance in molecular conformer generation, there are areas where it can be enhanced. One such area is the recall metrics, which capture the diversity of generated conformations. Another area is the use of an additional chirality correction step that is used to predict conformations with the desired chirality. Moreover, although our performance on the GEOM-XL dataset is comparable to MCF-S and TorsionDiff, there is still room for improvement.

We propose three future directions here. First, we observe during experiments that a well-designed sampling process incorporating stochasticity can enhance the quality and diversity of generated samples. An extension of our current approach could involve using Stochastic Differential Equations (SDEs), which utilize both vector field and score in the integration process, potentially improving the diversity of samples. Second, we propose to scale the number of parameters of {\method}, which has not only been shown to be useful across different domains of deep learning, but has also shown to be useful in molecular conformer generation for MCF \citep{wang2024swallowing}. Third, to better handle the chirality problem, we aim to explore alternatives for incorporating \textit{SO(3)}-equivariance into the model in the future.

 % instead of an additional chirality correction step, we aim to incorporate this directly into the architecture in the future
% \citep{ma2024sit} show that the score $s(t, \x)$ can be derived from the vector field $v(t, \x)$ for a one-sided interpolation with a Gaussian distribution without having to learn the separate score-matching model. This formulation can be generalized to a probability path between any two arbitrary distributions.

\subsubsection*{Acknowledgements}
The authors sincerely thank Cristian Gabellini, Jiarui Ding, and the NeurIPS reviewers for the insightful discussions and feedback. Resources used in completing this research were provided by Valence Labs. Furthermore, we acknowledge a grant for student supervision received by Mila - Quebec's AI institute - and financed by the Quebec ministry of Economy.

% \newpage
% \usepackage[authoryear]{natbib}
\bibliographystyle{plainnat}

\bibliography{reference}

% Use unnumbered third level headings for the acknowledgments. All acknowledgments go at the end of the paper. Do not include acknowledgments in the anonymized submission, only in the final paper.

% \section*{References}
\small

% References follow the acknowledgments. Use unnumbered first-level heading for the references. Any choice of citation style is acceptable as long as you are consistent. It is permissible to reduce the font size to \verb+small+ (9 point) when listing the references. {\bf Remember that you can use more than eight pages as long as the additional pages contain \emph{only} cited references.}
% \medskip

% \small

\newpage
\appendix
\section{Implementation Details}\label{app:implementation}
\begin{figure*}[h]
    \centering
    \includegraphics[width=\textwidth]{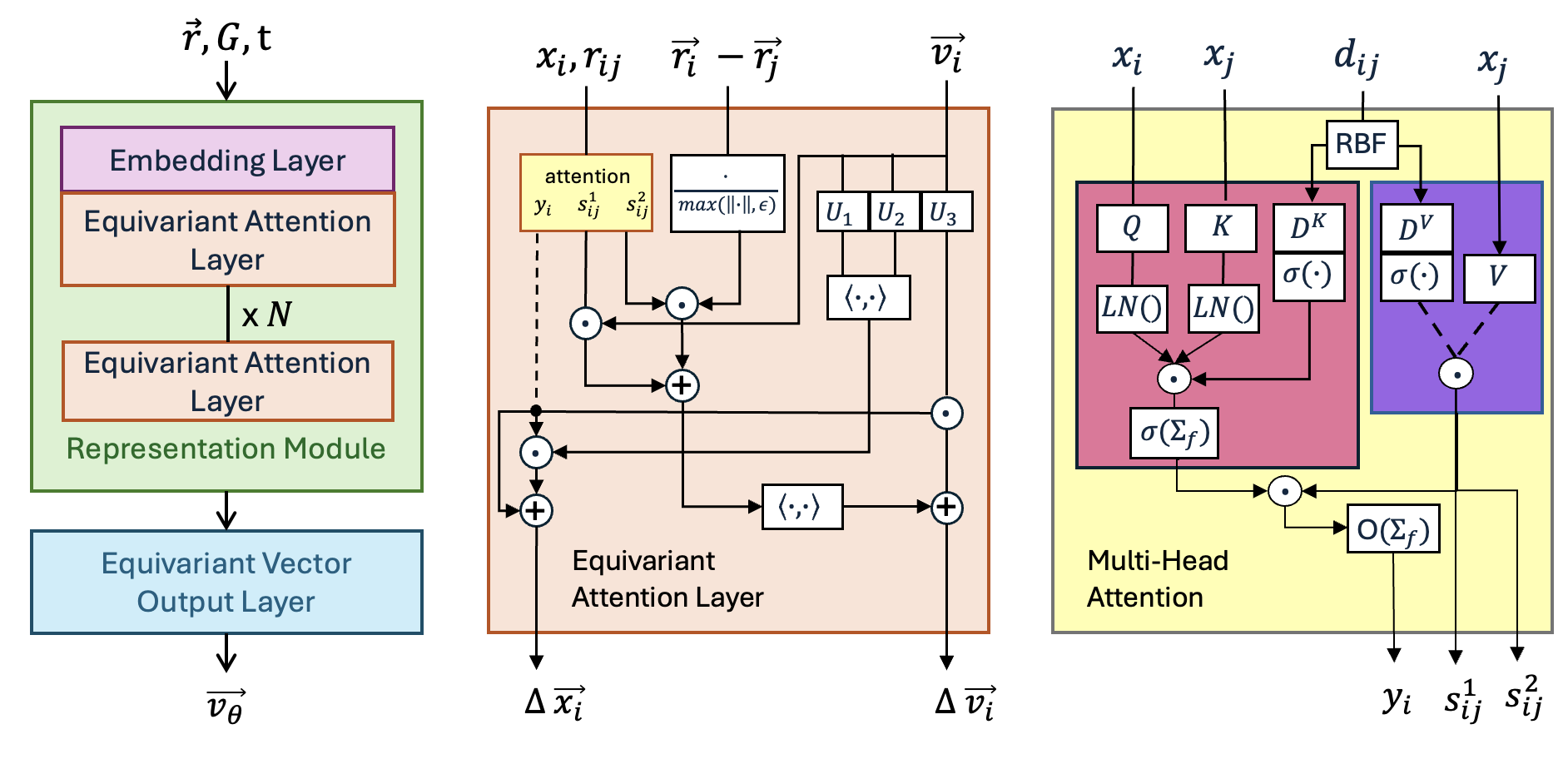}
    \caption{(a) Overall Architecture of {\method} consisting of 2 components, (1) Representation Layer based on TorchMD-NET \cite{tholke2022torchmd} and (2) Equivariant Output Layer from \citep{schutt2018schnet}. (b) Equivariant Attention Layer with all the operations involved, (c) Multi-Head Attention block modified with the LayerNorm.}
    \label{fig:architecture} 
\end{figure*}
\subsection{Architecture}
\label{app:architecture}
    The {\method} architecture (\autoref{fig:architecture}) consists of 2 major components, a representation layer and an output layer. For the representation layer, we use a modified version of the embedding and equivariant attention-based update layers from the equivariant transformer architecture of TorchMD-NET \citep{tholke2022torchmd}. The output layer utilizes the gated equivariant blocks from \citep{schutt2018schnet}. \md{We highlight our modifications over the original TorchMD-NET architecture with this color}. These modifications enable stabilized training since we use a larger network than the one proposed in the TorchMD-NET \citep{tholke2022torchmd} paper. Additionally, since our input structures are interpolations between structures sampled from a prior and actual conformations, it is important to ensure our network is numerically stable when the interpolations contain two atoms very close to each other. 

\textbf{Embedding Layer}: The embedding layer maps each atom's physical and chemical properties into a learned representation space, capturing both local atomic features and geometric neighborhood information. For the $i$-th atom in a molecule with $N$ atoms, we compute an invariant embedding $x_i$ through the following process:
\begin{align}
    z_i &= \text{embed}^{\text{int}}(z_i)  \\\
    h_i &= \text{MLP}(h_i)
\end{align}
where $z_i$ is the atomic number and $h_i$ represents atomic attributes (detailed in Appendix A). \md{The MLP projects atomic attributes into a feature vector of dimension $d_h$}.

Next, we compute a neighborhood embedding $n_i$ that captures local atomic environment:
\begin{align}
n_i &= \sum_{j=1}^{N} \text{embed}^{\text{nbh}}(z_j) \cdot g(d_{ij}, l_{ij}).
\end{align}
Here, $\text{embed}^{\text{nbh}}(z_j)$ provides a separate embedding for neighboring atomic numbers, $d_{ij}$ is the distance between atoms $i$ and $j$, and $l_{ij}$ encodes edge features (either from a radius-based graph or molecular bonds). The interaction function $g(d_{ij}, l_{ij})$ combines distance and edge information:
\begin{align}
g(d_{ij}, l_{ij}) &= W^F\left[\phi(d_{ij})e_1^{\text{RBF}}(d_{ij}), \ldots, \phi(d_{ij})e_K^{\text{RBF}}(d_{ij}), l_{ij}\right]
\end{align}
where $e_k^{\text{RBF}}$ are $K$ exponential radial basis functions following \citep{unke2019physnet}, and $\phi(d_{ij})$ is a smooth cutoff function:
\begin{align}
\phi(d_{ij}) &= \begin{cases}
\frac{1}{2}\left(\cos(\frac{\pi d_{ij}}{d_{\text{cutoff}}} + 1)\right), & \text{if } d_{ij} \leq d_{\text{cutoff}}\ \\
0, & \text{otherwise}
\end{cases}
\end{align}
Finally, we combine all features into the atom's embedding through a linear projection:
\begin{equation}
x_i = W^C \left[\text{embed}^{\text{int}}(z_i), h_i, t, n_i\right]
\end{equation}
where $t$ represents the time-step, and $[\cdot,\cdot]$ denotes concatenation. The resulting embedding $x_i \in \mathbb{R}^d$ serves as input to subsequent layers of the network.

\textbf{Attention Mechanism}: The multi-head dot-product attention operation uses atom features $x_i$, atom attributes $h_i$, time-step $t$ and inter-atomic distances $d_{ij}$ to compute attention weights. \md{The input atom-level features $x_i$ are mixed with the atom attributes $h_i$ and the time-step $t$ using an MLP and then normalized using a LayerNorm \citep{ba2016layer}}. To compute the attention matrix, the inter-atomic distances $d_{ij}$ are projected into two dimensional filters $D^K$ and $D^V$ as:
\begin{align}
    D^{K} &= \sigma \left(W^{D^K} e^{RBF}(d_{ij}) + b^{D^K}\right) \nonumber\\
    D^{V} &= \sigma \left(W^{D^V} e^{RBF}(d_{ij}) + b^{D^V}\right)
\end{align}
The atom level features are then linearly projected along with a LayerNorm operation to derive the query $Q$ and key $K$ vectors. The value vector $V$ is computed with only the linear projection of atom-level features. \md{Applying LayerNorm on Q, K vectors (also referred to as QK-Norm) has proven to stabilize un-normalized values in the attention matrix \citep{dehghani2023scaling, esser2024scaling} when scaling networks to large number of parameters.} The $Q$ and $K$ vectors are then used along with the distance filter $D^K$ for a dot-product operation over the feature dimension:
\begin{align}
    Q = \text{LayerNorm}(W^Q x_i), \quad K &= \text{LayerNorm}(W^K x_i),\quad V = W^V x_i\\
    \text{dot}(Q, K, D^K) &= \sum_{k}^F Q_k \cdot K_k \cdot D_k^K.
\end{align}
The attention matrix is derived by passing the above dot-product operation matrix through a non-linearity and weighting it using a cosine cutoff $\phi(d_{ij})$ (similar to the embedding layer) which ensures the attention weights are non-zero only when two atoms are within a specified cutoff:
\begin{align}
    A = \text{SiLU}(\text{dot}(Q, K, D^K)) \cdot \phi(d_{ij}).
\end{align}
Using the value vector $V$ and the distance filter $D_V$, we derive 3 equally sized filters by splitting along the feature dimension,
\begin{equation}\label{eq:s_split}
s_{ij}^1, s_{ij}^2, s_{ij}^3 = \text{split}(V_j \cdot D^V_{ij}).
\end{equation}
A linear projection is then applied to combine the attention matrix and the vectors $s_{ij}^3$ to derive an atom level feature $y_i = W^O\left(\sum_{j}^{N} A_{ij} \cdot s_{ij}^3\right)$. The output of the attention operation are $y_i$ (an atom level feature) and two scalar filters $s_{ij}^1$ and $s_{ij}^2$ (edge-level features).

\textbf{Update Layer}: The update layer computes interactions between atoms in the attention block and uses the outputs to update the scalar feature $x_i$ and the vector feature $\vec{v}_i$. First, the scalar feature output $y_i$ from the attention mechanism is split into three features ($q_i^1, q_i^2, q_i^3$), out of which $q_i^1$ and $q_i^2$ are used for the scalar feature update as,
\begin{equation}
    \Delta x_i = q_i^1 + q_i^2 \cdot \langle U_1\vec{v}_i\cdot U_2\vec{v}_i\rangle,
\end{equation}
where $\langle U_1\vec{v}_i\cdot U_2\vec{v}_i \rangle$ is the inner product between linear projections of vector features $\vec{v}_i$ with matrices $U_1, U_2$.

The edge vector update consists of two components. First, we compute a vector $\vec{w}i$, which for each atom is computed as a weighted sum of vector features and \md{a clamped-norm of the edge vectors} over all neighbors:
\begin{equation}
\label{eq:update_vec}
    \vec{w}_i = \sum_{j}^N s_{ij}^1 \cdot \vec{v}_j + s_{ij}^2 \cdot \frac{\vec{r}_i - \vec{r}_j}{\max(\lVert \vec{r}_i - \vec{r}_j\rVert, \epsilon)},
    % \vec{w}_i &= \sum_{j}^N s_{ij}^1 \cdot \vec{v}_j + s_{ij}^2 \cdot \frac{\vec{r}_i - \vec{r}_j}{\max(\lVert \vec{r}_i - \vec{r}_j\rVert, \epsilon)} + \add{s_{ij}^3 \cdot \left(\vec{v}_j \times \frac{\vec{r}_i - \vec{r}_j}{\max(\lVert \vec{r}_i - \vec{r}_j\rVert, \epsilon}\right)},\\
\end{equation}
\begin{equation}
    \Delta \vec{v}_i = \vec{w}_i + q_i^3 \cdot U_3 \vec{v}_i\\
    % \Delta \vec{v}_i &= \vec{w}_i + q_i^3 \cdot U_3 \vec{v}_i + \md{q_i^4 \langle U_3 \vec{v}_i \times U_1 \vec{v}_i \rangle}
\end{equation}
where $U_1$ and $U_3$ are projection matrices over the feature dimension of the vector feature $\vec{v}_i$. In this layer, \md{we clamp the minimum value of the norm (to $\epsilon = 0.01$) to prevent numerically large values in cases where positions of two atoms are sampled too close from the prior.}

\textbf{$SO(3)$ Update Layer}: \md{We also design an $SO(3)$ equivariant architecture by adding an additional cross product term in \autoref{eq:update_vec} as follows,}
\begin{equation}
    \vec{w}_i = \sum_{j}^N s_{ij}^1 \cdot \vec{v}_j + s_{ij}^2 \cdot \frac{\vec{r}_i - \vec{r}_j}{\max(\lVert \vec{r}_i - \vec{r}_j\rVert, \epsilon)} + s_{ij}^4 \cdot \left(\vec{v}_j \times \frac{\vec{r}_i - \vec{r}_j}{\max(\lVert \vec{r}_i - \vec{r}_j\rVert, \epsilon}\right),
\end{equation}
where $s{ij}^4$ is derived by modifying the split operation \autoref{eq:s_split} in the attention layer where the value vector $V$ and distance filter $D_V$ is projected into 4 equally sized filters instead of 3.

\textbf{Output Layer}: The output layer consists of Gated Equivariant Blocks from \citep{schutt2018schnet}. Given atom scalar $x_i$ and vector features $\vec{v}_i$, the updates in each block is defined as,
\begin{align}
    x_{i, \text{updated}}, \vec{w}_i &= \text{split}(\text{MLP}([x_i, U_1\vec{v}_i)])) \\
    \vec{v}_{i, \text{updated}} &= (U_2\vec{v}_i) \cdot \vec{w}_i
\end{align}
Here, $U_1$ and $U_2$ are linear projection matrices that act along feature dimension. \md{Our modification is to use LayerNorm in the MLP to improve training stability.}

\subsection{Input Featurization}
\label{app:atomic}

Atomic features (or Node Features) are computed using RDKit \citep{landrum2013rdkit} features as described in \autoref{tab:atomic}. For computing edge features and edge index, we use a combination of global (radius based edges) and local (molecular graph edges) similar to \citep{jing2022torsional}.

\begin{table*}[ht]
  \centering
  \small
  \begin{tabular}{l l l}
    \toprule
    Name & Description & Range \\
    \midrule
    \texttt{chirality} & Chirality Tag & \{unspecified, tetrahedral CW \& CCW, other\} \\
    \texttt{degree} & Number of bonded neighbors & $\{x:0 \leq x \leq 10, x \in \mathbb{Z}\}$ \\
    \texttt{charge} & Formal charge of atom & $\{x:-5 \leq x \leq 5, x \in \mathbb{Z}\}$ \\
    \texttt{num\_H} & Total Number of Hydrogens & $\{x:0 \leq x \leq 8, x \in \mathbb{Z}\}$ \\
    \texttt{number\_radical\_e} & Number of Radical Electrons & $\{x:0 \leq x \leq 4, x \in \mathbb{Z}\}$ \\    
    \texttt{hybrization} & Hybrization type & \{sp, sp\textsuperscript{2}, sp\textsuperscript{3}, sp\textsuperscript{3}d, sp\textsuperscript{3}d\textsuperscript{2}, other\} \\
    \texttt{aromatic} & Whether on a aromatic ring & \{True, False\} \\
    \texttt{in\_ring} & Whether in a ring & \{True, False\} \\
    \bottomrule
  \end{tabular}
  \caption{Atomic features included in {\method}.}
  \label{tab:atomic}
\end{table*}

% \subsection{Improving Stability of Training}
% We notice that when models receive unphysical inputs such as structures sampled from a Gaussian, atoms might end up too close together causing large gradients during training and issues with sampling. In order to alleviate these issues and to ensure smoother training, we modify the architecture with three changes: (1) additional layer norm in the update layer of the equivariant output layer that acts on invariant tensors, (2) clamped normalization of edge vectors in the equivariant transformer backbone, where we clamp the minimum value of the norm to $0.01$ and (3) usage of QK-norm \citep{dehghani2023scaling, esser2024scaling} where a LayerNorm is applied on the keys and queries before the attention operation.

\subsection{Evaluation Metrics}
Following the approaches of \citep{ganea2021geomol, xu2022geodiff, jing2022torsional}, we utilize Average Minimum RMSD (AMR) and Coverage (COV) to assess the performance of molecular conformer generation. Here, $C_g$ denotes the set of generated conformations, and $C_r$ denotes the set of reference conformations. For both AMR and COV, we calculate and report Recall (R) and Precision (P). Recall measures the extent to which the generated conformers capture the ground-truth conformers, while Precision indicates the proportion of generated conformers that are accurate. The specific formulations for these metrics are detailed in the following equations:
\begin{equation*}
    \text{AMR-R}(C_g, C_r) = \frac{1}{|C_r|} \sum_{\mathbf{R} \in C_r} \min_{\mathbf{\hat{R}} \in C_g} \text{RMSD}(\mathbf{R}, \mathbf{\hat{R}})
\label{eq:amrr}
\end{equation*}
\begin{equation*}
    \text{COV-R}(C_g, C_r) = \frac{1}{|C_r|} |\{ \mathbf{R} \in C_r | \text{RMSD}(\mathbf{R}, \mathbf{\hat{R}}) < \delta, \mathbf{\hat{R}} \in C_g \}|
\label{eq:covr}
\end{equation*}

\begin{equation*}
    \text{AMR-P}(C_r, C_g) = \frac{1}{|C_g|} \sum_{\mathbf{\hat{R}} \in C_g} \min_{\mathbf{R} \in C_r} \text{RMSD}(\mathbf{\hat{R}}, \mathbf{R})
\label{eq:amrp}
\end{equation*}
\begin{equation*}
    \text{COV-P}(C_r, C_g) = \frac{1}{|C_g|} |\{ \mathbf{\hat{R}} \in C_g | \text{RMSD}(\mathbf{\hat{R}}, \mathbf{R}) < \delta, \mathbf{R} \in C_r \}|
\label{eq:covp}
\end{equation*}

A lower AMR score signifies improved accuracy, while a higher COV score reflects greater diversity in the generative model. Following \citep{jing2022torsional}, the threshold $\delta$ is set to 0.5$\text{\AA}$ for GEOM-QM9 and 0.75$\text{\AA}$ for GEOM-DRUGS.

\subsection{Training Details and Hyperparameters}\label{app:hyperparameters}

\begin{table*}[h]
\scriptsize
    \centering
    \footnotesize
    \setlength{\tabcolsep}{10pt}
    \begin{tabular}{l c}
    \toprule
        Hyper-parameter & {\method} \\
        \midrule
        % \texttt{num\_eigenfuncs} & $28$ & $32$ \\
        \texttt{num\_layers}                & $20$  \\
        \texttt{hidden\_channels}           & $160$  \\ 
        \texttt{num\_heads}                 & $8$    \\
        \texttt{neighbor\_embedding}        & True   \\
        \texttt{cutoff\_lower}              & $0.0$  \\
        \texttt{cutoff\_higher}             & $10.0$ \\
        \texttt{node\_attr\_dim}            & $8$    \\
        \texttt{edge\_attr\_dim}            & $1$    \\
        \texttt{reduce\_op}                 & True  \\
        \texttt{activation}                 & SiLU  \\
        \texttt{attn\_activation}           & SiLU  \\
        \midrule
        \texttt{\# param}                   & 8.3M \\
        \bottomrule
    \end{tabular}
    \caption{Hyperparameters for {\method}}
    \label{tab:et-flow-hparams}
\end{table*}

For GEOM-DRUGS, we train {\method} for a fixed $250$ epochs with a batch size of $64$ and $5000$ training batches per epoch per GPU on 8 A100 GPUs. For the learning rate, we use the Adam Optimizer with a cosine annealing learning rate which goes from a maximum of $10^{-3}$ to a minimum $10^{-7}$ over 250 epochs with a weight decay of $10^{-10}$. For GEOM-QM9, we train {\method} for $200$ epochs with a batch size of 128, and use all of the training dataset per epoch on $4$ A100 GPUs. We use the cosine annealing learning rate schedule with maximum of $8 \cdot 10^{-4}$ to minimum of $10^{-7}$ over 100 epochs, post which the maximum is reduced by a factor of $0.05$. We select checkpoints based on the lowest validation error. %For ablations, experiments are conducted using models trained on $2$ GPUs for $100$ epochs on GEOM-QM9. The hyperparameters for the experiments are shared in \autoref{tab:et-flow-hparams}.

\section{Training and Sampling Algorithm} \label{app:summary_procedures}

The following algorithms go over the pseudo-code for the training and sampling procedure. For each molecule, we use up to $30$ conformations with the highest boltzmann weights as provided by CREST \citep{pracht2024crest} similar to that of \citep{jing2022torsional}

\begin{algorithm}
\caption{Training procedure}\label{alg:training}
\KwIn{molecules $[G_0, ..., G_N]$ each with true conformers $[C_{G,1}, ... C_{G,K_G}]$, the harmonic prior $\rho_0$, learning rate $\alpha$, number of epochs $N_e$, initialized vector field $v_\theta$}

\KwOut{trained flow matching model $v_\theta$}
% conformer matching process for each $G$ to get $[\hat{C}_{G,1}, ... \hat{C}_{G,K_G}]$\;
\For{$i \leftarrow 1$ \KwTo $N_e$}{
    \For{$G$ \textbf{in} $[G_0, ..., G_N]$}{
        Sample $t \sim \U[0, 1]$ and $C_1 \in [C_{G,1}, ...C_{G,K_G}]$\;
        Sample prior $C_0 \sim \rho_0(G)$\;
        Align $C_0 \leftarrow \text{RMSDAlign}(C_0, C_1)$\;
        % Sample $C_t \sim \N\left(tC_1 + (1 - t)C_0, \sigma^2 t(1 - t)\right)$\;
        Sample $C_t = tC_1 + (1 - t)C_0 + \sigma^2 t(1 - t)z, \hspace{0.5 cm} z \sim \N(0, \I)$\;
        Construct vector field $u_t \leftarrow x_1 - x_0 + \frac{1 - 2t}{2 \sqrt{t (1 - t)}} z$\;
        Compute loss $\mathcal{L} \leftarrow \| v_\theta(t, C_t) - u_t \|^2 $\;
        Take gradient step $\theta \leftarrow \theta - \alpha \nabla_{\theta}\mathcal{L}$\;
    }
}
\end{algorithm}

\begin{algorithm}[h]
\caption{Inference procedure}\label{alg:inference}
\KwIn{molecular graph $G$, number conformers $K$, number of sampling steps $N$}
\KwOut{predicted conformers  $[C_1, ... C_K]$}
\For{$C$ \textbf{in} $[C_1, ... C_K]$}{
    sample prior $\hat{C} \sim \rho_0(G)$\;
    \For{n $\leftarrow 0$ \KwTo $N - 1$}{
        Set $t \leftarrow \frac{n}{N}$\;
        Set $\Delta t \leftarrow \frac{1}{N}$\;
        Predict $\hat{v} = v_\theta(t, \hat{C})$\;
        Update $\hat{C} = \hat{C} + \hat{v}\Delta t$
    }
}
\end{algorithm}

\begin{algorithm}[h!]
\caption{Stochastic Sampler}\label{alg:stochastic}
\KwIn{molecular graph $G$, number conformers $K$, number of sampling steps $N$, stochasticity level $churn$, stochastic sampling range $[t_{min}, t_{max}]$ }
\KwOut{predicted conformers  $[C_1, ... C_K]$}
\For{$C$ \textbf{in} $[C_1, ... C_K]$}{
    sample prior $\hat{C} \sim \rho_0(G)$\;
    \For{n $\leftarrow 0$ \KwTo $N - 1$}{
        Set $t \leftarrow \frac{n}{N}$\;
        Set $\Delta t \leftarrow \frac{1}{N}$\;
        Set $\gamma \leftarrow \frac{churn}{N}$\;
        \If {$t \in [t_{min}, t_{max}]$} {
            Sample $\epsilon \sim N(0, I)$\;
            $\Delta \hat{t} \leftarrow \gamma (1-t)$\;
            $\hat{t} \leftarrow max(t - \Delta \hat{t}, 0)$\;
            $\hat{C} \leftarrow \hat{C} + \Delta \hat{t} \sqrt{t^2-\hat{t}^2}\epsilon$\;
            Predict $\hat{v} = v_\theta(\hat{t}, \hat{C})$\;
            Set $\Delta t \leftarrow \Delta t + \Delta \hat{t}$\;
            }
        \Else {
        Predict $\hat{v} = v_\theta(t, \hat{C})$\;
        }
        Update $\hat{C} = \hat{C} + \hat{v}\Delta t$
    }
}
\end{algorithm}\label{alg:sto-sampler}

\newpage

\section{Proofs}
\subsection{Designing SO(3) Equivariance}\label{app:proof_of_so3}
We show that we can modify the architecture in \autoref{app:architecture} (Equation 18) to produce a final vector output that satisfies rotation equivariance and reflection asymmetry. Let $\vv_1$ and $\vv_2$ be linearly independent non-zero vectors $\norm{\vv_1}>0, \norm{\vv_2}>0$, and $s$ be a scalar. We implement SO(3) equivariance by adding a vector with a cross product. We show that vector $\vv = \vv_1 + s (\vv_1 \times \vv_2)$, where $\vv_1 \times \vv_2$ denotes cross product of $\vv_1$ and $\vv_2$, satisfies anti-symmetry while maintaining rotation equivariance as follows,

\begin{align}
    R\vv_1 + s(R\vv_1 \times R\vv_2)   &= R(\vv_1) + s R(\vv_1 \times \vv_2) \\
                                    &= R( \vv_1 + s (\vv_1 \times \vv_2) ) \\ 
    -\vv_1 + s(-\vv_1 \times -\vv_2)   &= -\vv_1 + s (\vv_1 \times \vv_2) \\
                                    &\neq -(\vv_1 + s (\vv_1 \times \vv_2)) 
\end{align}
This concludes the proof for rotation equivariance and reflection anti-symmetry.

\section{Additional Results} \label{app:additional_results}

\subsection{Design Choice Ablations}\label{sec:ablation}
We conduct a series of ablation studies to assess the influence of each component in the {\method}. Particularly, we re-run the experiments with (1) $O(3)$ equivariance without chirality correction, (2) Absence of Alignment, (3) Gaussian Prior as a base distribution. We demonstrate that improving probability paths and utilizing an expressive equivariant architecture with correct symmetries are key components for {\method} to achieve state of the art performance. The ablations were ran with reduced settings ($50$ epochs; lr $=1e-4$; $4$ A100 gpus). Results are shown in \autoref{tab:ablations}.
% , (4) Using a less expressive equivariant architecture, EGNN \cite{satorras2021n} as a model backbone.

\begin{table*}[ht]
\centering
\caption{Ablation results on GEOM-DRUGS.}
\vskip 0.1in
\begin{adjustbox}{max width=\textwidth}
\renewcommand{\arraystretch}{1.1} % Default value: 1
\begin{tabular}{lcccccccc}
\midrule
 & \multicolumn{4}{c}{Recall} & \multicolumn{4}{c}{Precision} \\
\Xhline{2\arrayrulewidth}
 & \multicolumn{2}{c}{Coverage $\uparrow$} & \multicolumn{2}{c}{AMR $\downarrow$} & \multicolumn{2}{c}{Coverage $\uparrow$} & \multicolumn{2}{c}{AMR $\downarrow$} \\
 \hline
 & mean & median & mean & median & mean & median & mean & median\\
    \hline
    {\method}                     & 75.37 & 82.35 & 0.557 & 0.529 & 58.90 & 60.87 & 0.742 & 0.690 \\
    {\method} ($O(3)$)            & 72.74 & 79.21 & 0.576 & 0.556 & 54.84 & 54.11  & 0.794 & 0.739 \\
    {\method} (w/o Alignment)     & 68.67 & 74.71 & 0.622 & 0.611 & 47.09 & 44.25 & 0.870 & 0.832 \\
    {\method} (Gaussian Prior)    & 66.53 & 73.01 & 0.640 & 0.625 & 44.41 & 40.88 & 0.903 & 0.864 \\
    % {\method} (EGNN)              & 0.613 & 0. & 1.569 & 1.561 & 0.27 & 0. & 2.077 & 2.071 \\    
%\Xhline{3\arrayrulewidth}
\midrule
\end{tabular}
\end{adjustbox}
\end{table*}\label{tab:ablations}

\subsection{Results on GEOM-XL}\label{sec:xl}
\begin{table}[ht]
\caption{Generalization results on GEOM-XL.}
\setlength{\tabcolsep}{3pt}
    \centering
    \small
\begin{tabular}{l cc cc c}
& \multicolumn{2}{c}{AMR-P $\downarrow$} & \multicolumn{2}{c}{AMR-R $\downarrow$} & \# mols \\   
\toprule
& mean & median & mean & median &   \\
\midrule
GeoDiff & 2.92 & 2.62 & 3.35 & 3.15 & - \\
GeoMol & 2.47 & 2.39 & 3.30 & 3.14 & - \\
Tor. Diff. & 2.05 & 1.86 & \textbf{2.94} & 2.78 & - \\
% {\model} (ours)  & 2.23 & 1.92 & 3.26 & 2.89 & 102 \\
MCF - S & 2.22 & 1.97 & 3.17 & 2.81 & 102 \\
MCF - B & 2.01 & 1.70 & 3.03 & 2.64 & 102 \\
MCF - L & \textbf{1.97} & \textbf{1.60} & \textbf{2.94} & \textbf{2.43} & 102 \\
{\method} (ours) & 2.31 & 1.93 & 3.31 & 2.84 & 102 \\
\midrule
Tor. Diff.   & 1.93  &	1.86   &	2.84 &	2.71 &	77 \\
% {\method} (ours)   &	1.97  &	\textbf{1.85}  &	2.98  &	2.72	& 77 \\
MCF - S & 2.02 & 1.87 & 2.9 & 2.69  & 77 \\
MCF - B & 1.71 & 1.61 & 2.69 & 2.44  & 77 \\
MCF - L & \textbf{1.64} & \textbf{1.51} & \textbf{2.57} & \textbf{2.26}  & 77 \\
{\method} (ours) & 2.00 & 1.80 & 2.96 & 2.63 & 75 \\
\bottomrule
\end{tabular}
\end{table}\label{tab:xl}
We now assess how well a model trained on GEOM-DRUGS generalises to unseen molecules with large numbers of atoms, using the GEOM-XL dataset containing a total of 102 molecules. This provides insights into the model's capacity to tackle larger molecules and out-of-distribution tasks. Upon executing the checkpoint provided by Torsional Diffusion, we encountered 27 failed cases for generation likely due to RDKit failures, similar to the observations in MCF albeit with slightly different exact numbers. In both experiments involving all 102 molecules and a subset of 75 molecules, {\method} achieves performance comparable to Torsional Diffusion and MCF-S, but falls short of matching the performance of MCF-B and MCF-L. It's worth noting that MCF-B and MCF-L are significantly larger models, potentially affording them an advantage in generalization tasks. As part of our future work, we plan to scale up our model and conduct further tests to explore its performance in this regard. %\ns{check how they defended in the workshop paper https://openreview.net/pdf?id=Od1KtMeAYo}

\subsection{Additional Out-of-Distribution Results}
% To assess the generalization performance, we evaluate {\method} on two out-of-distribution experiments. First, we assess the model scaffold-based splits rather than the standard random split on the GEOM-QM9 dataset. \mh{insert details of the split}. The second experiment takes a model trained on GEOM-DRUGS and evaluates its performance on GEOM-QM9, which contains significantly smaller molecules. This complements the experiment on generalizing to larger molecules in GEOM-XL. The results can be found in \autoref{tab:ood-results}. We observe that the model performance only degrades marginally on the scaffold-based split. We show that our model still has fairly robust performance on GEOM-QM9 when trained on GEOM-DRUGS.

\begin{table*}[ht] % Changed from table* to table
\centering
\caption{Additional OOD results. We use RS and SS to indicate Random Split and Scaffold Split respectively.}
\vskip 0.1in
\begin{adjustbox}{max width=\textwidth}
\renewcommand{\arraystretch}{1.1} % Default value: 1
\begin{tabular}{lcccccccc} % Reduced width and added horizontal padding
\toprule
 & \multicolumn{4}{c}{Recall} & \multicolumn{4}{c}{Precision} \\
\Xhline{2\arrayrulewidth}
 & \multicolumn{2}{c}{Coverage $\uparrow$} & \multicolumn{2}{c}{AMR $\downarrow$} & \multicolumn{2}{c}{Coverage $\uparrow$} & \multicolumn{2}{c}{AMR $\downarrow$} \\
 \hline
 & mean & median & mean & median & mean & median & mean & median\\
    \hline
    {\method} (QM9 RS)                     &  96.47 & 100.00 & 0.073 & 0.047 & 94.05 & 100.00 & 0.098 & 0.039 \\
    {\method} (QM9 SS)                  & 95.00 & 100.00 & 0.083 & 0.029 & 90.25 & 100.00 & 0.124 & 0.053 \\
    {\method} (DRUGS $\rightarrow$ QM9)                  & 86.68 & 100.00 & 0.218 & 0.160 & 68.69 & 75.30  & 0.369 & 0.317 \\
    \midrule
    {\method} (DRUGS RS)      &  79.53 & 84.57 & 0.452 & 0.419 & 74.38 & 81.04 & 0.541 & 0.470 \\ %post-hoc correction 
    {\method} (DRUGS SS)                  & 76.06 & 80.65 & 0.644 & 0.545 & 67.83 & 74.19 & 0.511 & 0.473 \\
    \bottomrule
\end{tabular}
\end{adjustbox}
\label{tab:ood-results}
\end{table*}

To further evaluate the generalization performance of {\method}, we conduct two more out-of-distribution experiments in addition to GEOM-XL. First, we test the model on scaffold-based splits of the GEOM-QM9 \add{and GEOM-DRUGS} dataset, which offers a more challenging alternative to the standard random split. We split the datasets based on Murcko scaffolds of the molecules into an 80:10:10 ratio for train, validation, and test sets. We evaluate our method on 1000 randomly sampled molecules from the resulting test set. The second experiment involves training the model on GEOM-DRUGS and assessing its performance on GEOM-QM9, a dataset with significantly smaller molecules. This experiment complements the generalization task to larger molecules in GEOM-XL by assessing ability for {\method} to generalize to smaller molecules. The results, presented in \autoref{tab:ood-results}, indicate that the model's performance degrades only marginally on the scaffold-based split. Furthermore, the model demonstrates robust performance on GEOM-QM9 even when trained on GEOM-DRUGS.

\section{Visualizations}

\begin{figure}[h]
    \centering
    \includegraphics[width=\columnwidth]{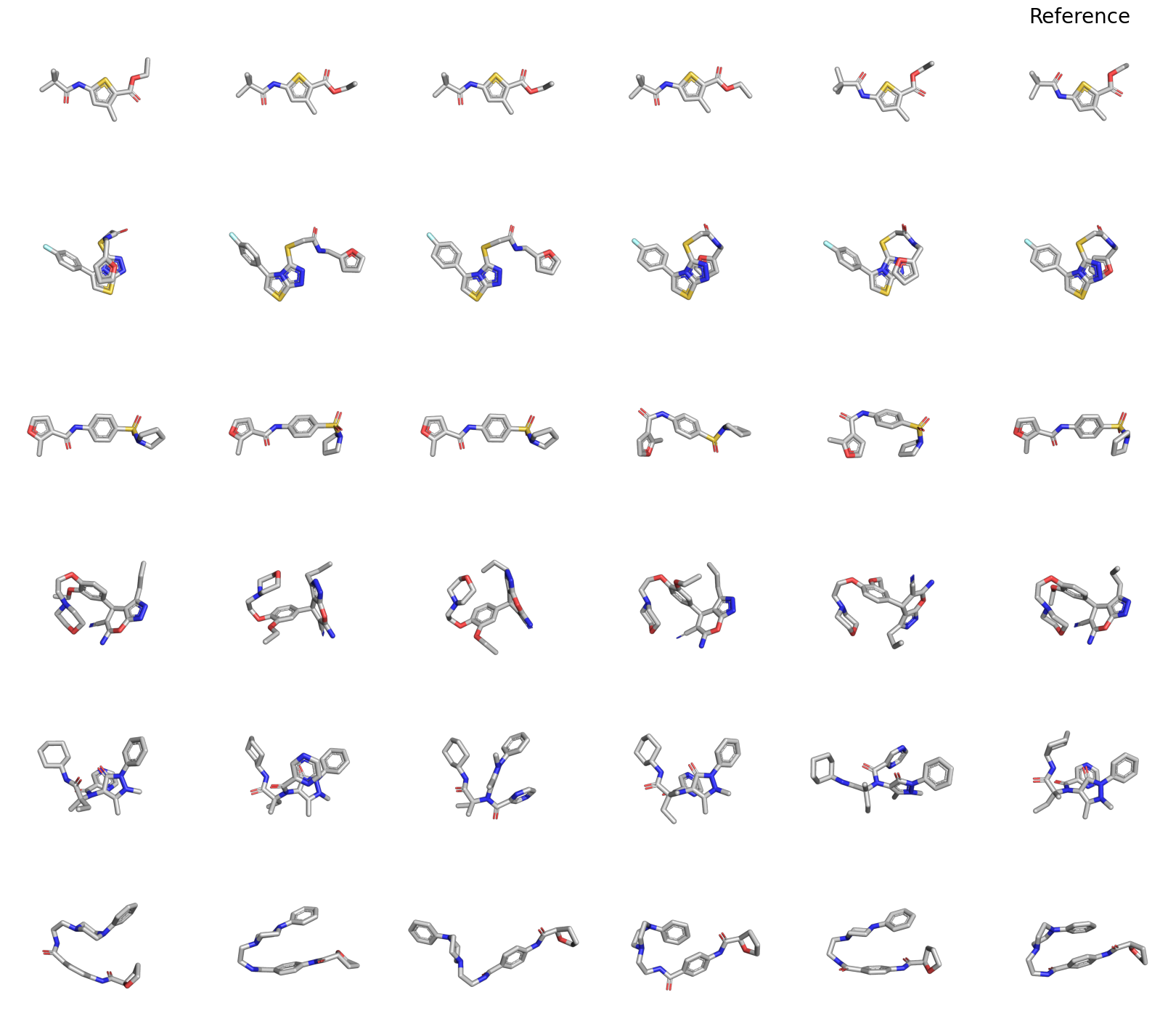}
    \caption{Examples of conformers of ground truth and {\method} for GEOM-DRUGS.}
    \label{fig:gen_mol}
\end{figure}

\begin{figure}[h]
    \centering
    \includegraphics[width=\columnwidth]{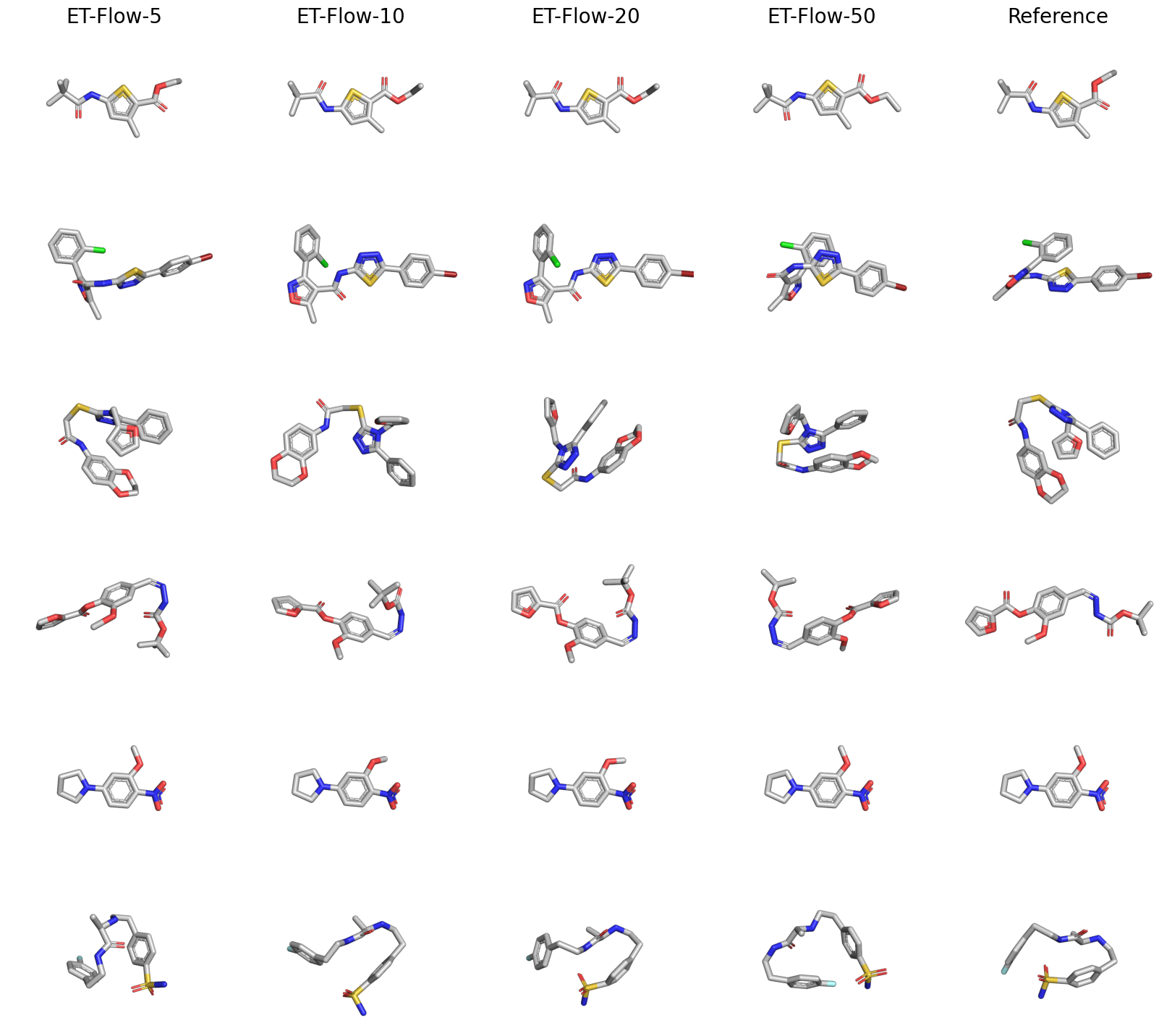}
    \caption{Examples of conformers of ground truth and {\method} for different number of sampling steps.}
    \label{fig:gen_mol_by_step}
\end{figure}

\autoref{fig:gen_mol} shows randomly selected examples of sampled conformers from {\method} for GEOM-DRUGS. The left column is the reference molecule from the ground truth, and the remaining columns are samples generated with $50$ sampling steps. \autoref{fig:gen_mol_by_step} showcases the ability for {\method} to generate quality samples with fewer sampling steps.

\clearpage

\end{document}

%% file: preamble.tex
    \PassOptionsToPackage{authoryear}{natbib}
\usepackage[final]{neurips_2024}

\usepackage[utf8]{inputenc} % allow utf-8 input
\usepackage[T1]{fontenc}    % use 8-bit T1 fonts
\usepackage[ruled,vlined]{algorithm2e}
\usepackage{capt-of} % for \captionof command
\usepackage{url}            % simple URL typesetting
\usepackage{booktabs}       % professional-quality tables
\usepackage{graphicx}
\usepackage{siunitx}
\usepackage{adjustbox}
\usepackage{amsfonts}       % blackboard math symbols
\usepackage{nicefrac}       % compact symbols for 1/2, etc.
\usepackage{microtype}      % microtypography
\usepackage{bm}
\usepackage{amsmath}
\usepackage{xcolor}
\usepackage{makecell}
\usepackage{subcaption}
\usepackage{wrapfig}

\docsvlist{R, E}

\docsvlist{L, U, N}

\docsvlist{x, z, I}

\newcommand{\vv}{\vec{v}}

\setcitestyle{authoryear, open={((},close={)}}
\definecolor{mydarkblue}{rgb}{0,0.2,0.60}
\definecolor{babyblue}{rgb}{0.54, 0.81, 0.94}
\definecolor{dark2orange}{RGB}{237, 127, 49}
\definecolor{awesome}{RGB}{191, 30, 19}
\definecolor{linkblue}{RGB}{30, 100, 230}

\usepackage[colorlinks=true, linkcolor=linkblue, urlcolor=black, citecolor=mydarkblue]{hyperref}

\DeclareSIUnit\angstrom{\text {Å}}
\newcommand{\md}[1]{{\color{awesome} #1}}

\newcommand{\add}[1]{{\color{blue} #1}}
\newcommand\norm[1]{\left\lVert#1\right\rVert}

\newcommand{\method}{ET-Flow}

\AtBeginDocument{}
\AtBeginDocument{}
\AtBeginDocument{}
\AtBeginDocument{}

\newcommand\nnfootnote[1]{%
  \renewcommand\thefootnote{}\footnote{#1}%
  \addtocounter{footnote}{-1}%
}